\documentclass[prl,twocolumn,aps,showpacs]{revtex4}
\usepackage{epsfig}
\usepackage{bm}
\begin{document}
\date{\today}

\title{The Josephson effect throughout the BCS-BEC crossover}
\author{A. Spuntarelli, P. Pieri, and G.C. Strinati}
\affiliation{Dipartimento di Fisica, 
Universit\`{a} di Camerino, I-62032 Camerino, Italy}

\begin{abstract}
We study the stationary Josephson effect for neutral fermions across the 
BCS-BEC crossover, by solving numerically the Bogoliubov-de Gennes equations at zero temperature.
The Josephson current is found to be considerably enhanced for all barriers at about unitarity.
For vanishing barrier, the Josephson critical current approaches the Landau limiting value which, depending on 
the coupling, is determined by either pair-breaking or sound-mode excitations. 
In the coupling range from the BCS limit to unitarity, a procedure is proposed to extract the pairing gap from the Landau limiting current.
\end{abstract}

\maketitle

There has been much interest lately in the BCS-BEC crossover, motivated by its experimental realization with ultracold trapped Fermi atoms.
Properties that have been measured include radio-frequency spectra aimed at extracting the pairing gap \cite{Grimm-2004-I,Jin-2005-I,Ketterle-2007} 
and vortices under rotation \cite{Ketterle-2005-II}.
In superconductors, these properties correspond to the occurrence of the gap parameter and the partial Meissner effect, which reveal the superconducting state.

The two limiting situations, of fermions described by BCS theory in weak coupling and of composite bosons undergoing Bose-Einstein condensation in strong coupling, are connected with continuity because they correspond to the same spontaneous breaking of the gauge symmetry associated with the phase of the complex order parameter.
In superconductors, this symmetry breaking is embodied by the Josephson effect, whereby a finite phase difference between two superconductors coupled via a link results in a steady current flow \cite{Josephson-1969}.

Bosonic Josephson junctions have recently been realized with Bose-Einstein condensates~\cite{boson-Josephson}.  
One can therefore foresee an experimental study of the Josephson effect also with ultracold trapped Fermi atoms, whose mutual attraction can be varied from weak to strong coupling with the use of a Fano-Feshbach resonance \cite{FF-exp}.
In this context, the recent experimental study of stationary vs non-stationary flow of a bosonic condensate through a potential barrier \cite{EA-2007} may serve as a guide for a corresponding experiment with ultracold Fermi atoms, aimed at measuring specifically the Josephson effect.

No theoretical study of the Josephson effect throughout the BCS-BEC crossover exists thus far, following the evolution between its Fermi and Bose versions.
Purpose of the present work is to fill at least partially this gap, by calculating the Josephson current/phase relation throughout the BCS-BEC crossover at 
zero temperature within a mean-field approach for an inhomogeneous system in the presence of a barrier.
The nontrivial results obtained by this study are, however, expected to hold even beyond mean field.
These results regard specifically the behavior of the Josephson characteristics and the maximum Josephson current for varying couplings and barriers, as well the relation of this maximum current to the Landau criterion for superfluidity across the BCS-BEC crossover.
Our study will, in fact, reveal an intimate connection between the maximum Josephson and Landau critical currents throughout the crossover.

We consider a system of neutral fermions, mutually interacting via an attractive short-range potential parametrized in terms of the scattering length $a_{F}$.
The BCS-BEC crossover is thus driven by the dimensionless coupling parameter
$(k_{F} a_{F})^{-1}$ where $k_{F}$ is the Fermi wave vector.
This parameter ranges from $(k_{F} a_{F})^{-1} \ll -1$ in the weak-coupling (BCS) limit to
$1 \ll (k_{F} a_{F})^{-1}$ in the strong-coupling (BEC) limit, while the crossover region of interest is restricted to 
$-1 \lesssim (k_{F} a_{F})^{-1} \lesssim +1$ with the
unitary limit at $(k_{F} a_{F})^{-1} = 0$.

To be specific, we realize the Josephson link with a \emph{slab} geometry, whereby a potential barrier 
$V(x)$ of width $L$ and height $V_{0} > 0$ is embedded in a homogeneous superfluid extending 
to infinity on both sides of the barrier.
Although the profile of the barrier is one-dimensional (with the $x$ direction orthogonal to the barrier), the slab is
fully three-dimensional as it extends along the two other ($y$ and $z$) directions parallel to the barrier.
This is because, on the BEC side of the crossover, the formation of composite bosons out of their 
fermionic constituents requires one to include wave vectors with components along all three dimensions. 
This marks a difference from previous treatments of the Josephson effect, that considered either fermions in the
weak-coupling limit only \cite{Wendin-1996} or point-like bosons whose formation is not an issue
\cite{Smerzi-1999}.
In addition, we regard the fermionic attraction to extend unmodified in the barrier region, thus anticipating
the situation with ultracold Fermi atoms.

To describe the effects of the potential barrier on the order parameter $\Delta(\mathbf{r})$  (where $\mathbf{r}=(x,y,z)$), we solve the Bogoliubov-de Gennes (B-dG) equations for the two-component fermionic wave functions \cite{BdG}:
\begin{equation}
\left( 
\begin{array}{cc}
\mathcal{H}(\mathbf{r}) & \Delta(\mathbf{r})            \\
\Delta(\mathbf{r})^{*}  & - \mathcal{H}(\mathbf{r})  
\end{array} 
\right)
\left( \begin{array}{c}
u_{\nu}(\mathbf{r}) \\
v_{\nu}(\mathbf{r}) 
\end{array} 
\right) 
= \epsilon_{\nu}
\left( \begin{array}{c}
u_{\nu}(\mathbf{r}) \\
v_{\nu}(\mathbf{r}) 
\end{array} 
\right) .                                    \label{B-dG-equations} 
\end{equation}
Here, $\mathcal{H}(\mathbf{r}) = -  \nabla^{2}/(2m) + V(x) - \mu$ where 
$m$ and $\mu$ are the fermion mass and chemical potential. 
The function $\Delta(\mathbf{r})$ is determined via the \emph{self-consistency condition}:
\begin{equation}
\Delta(\mathbf{r}) = g \sum_{\nu} u_{\nu}(\mathbf{r}) 
v_{\nu}(\mathbf{r})^{*}       \label{self-consistency}
\end{equation}
where $- g$ is the strength of the local fermionic attraction, eliminated eventually in favor of the scattering length 
$a_{F}$~\cite{deMelo93}.
These equations have previously been considered in the weak-coupling (BCS) limit, to determine the Josephson current for one-dimensional situations \cite{Wendin-1996}.

As for any approach to a crossover problem, the present study of the Josephson effect has also to rely on a definite benchmark in the strong-coupling (BEC) limit.
In this context, it has been shown~\cite{PS-2003} that the fermionic B-dG equations (\ref{B-dG-equations}) and (\ref{self-consistency}) can be suitably mapped
onto the Gross-Pitaevskii (GP) equation \cite{Gross,Pitaevskii}    
\begin{equation}
\left[-\frac{\nabla^{2}}{4 m} + 2 V(x)  +
\frac{8 \pi a_{F}}{ 2 m} |\Phi(\mathbf{r})|^{2} \right] \Phi(\mathbf{r}) = \mu_{B} \Phi(\mathbf{r}) 
 \label{Gross-Pitaevskii-equation}                  
\end{equation}
for the condensate wave function $\Phi(\mathbf{r})$ of composite bosons with mass $2m$, which form in the strong-coupling limit of the fermionic attraction.
In the above equation, $V(x)$ is the same potential of Eqs.(\ref{B-dG-equations}) and
$\mu_{B} = 2 \mu + \varepsilon_{0}$ is the chemical potential for composite bosons ($\varepsilon_{0}>0$ being the two-fermion binding energy such that 
$\mu_{B}\ll \varepsilon_{0}$).
The two functions $\Phi(\mathbf{r})$ of Eq.(\ref{Gross-Pitaevskii-equation}) and $\Delta(\mathbf{r})$ of Eq.(\ref{self-consistency}) are related by 
$\Phi(\mathbf{r}) = \Delta(\mathbf{r}) \sqrt{(m^{2} a_{F})/8 \pi}$.
This mapping has been established \cite{PS-2003} for low-enough temperatures, so that all composite bosons reside in the condensate.
This restriction matches one's expectation that a mean-field approach suffices to describe (at least qualitatively) the BCS-BEC crossover at low enough 
temperatures.  

The self-consistent solution of the B-dG equations first proceeds by extending to arbitrary values of the fermionic attraction the numerical approach 
introduced in Ref.\cite{Bagwell-1996} for weak coupling. 
The profile of $\Delta(x)$ is made piecewise constant over a dense number of intervals (typically $80$), in such a way that the scattering problem is 
solved by elementary methods within each interval for given fermionic quasi-particle energy, while the scattering wave functions in contiguous intervals are connected by 
continuity conditions.
Outgoing boundary conditions for waves impinging on the barrier from the left and right are used to identify the contribution of the continuos energies of 
Eqs.(\ref{B-dG-equations}).
The contribution of bound levels (known as Andreev-Saint James states \cite{Andreev-states}) needs also be included. Self-consistency for the gap profile is implemented from the outcomes of such a scattering problem over a less dense grid of points (typically $20$). 
Appropriate numerical strategies are required to speed up self-consistency, which is then reached with a limited number of cycles (typically $5$).  
                                                                                  
Numerical calculations have been performed by imposing \emph{either} the value $n_{0} q/m$ of the Josephson current (provided it can be self-consistently sustained) where
$n_{0}$ is the bulk density, \emph{or} the value of the asymptotic phase difference $\delta \phi=\phi(x=+\infty)-\phi(x=-\infty)$ between the two sides of the barrier.
Under these circumstances, we write $\Delta(x)=|\Delta(x)|\exp[ 2 i q x + i \phi(x)]$.     

\begin{figure}
\begin{center}
\epsfxsize=6.5cm
\epsfbox{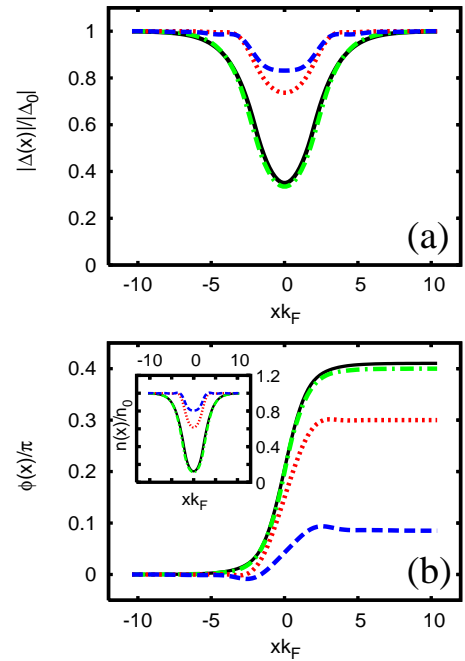}
\caption{(a) Magnitude $|\Delta(x)|$ and (b) phase $\phi(x)$ of the order parameter vs the coordinate $x$ orthogonal to the barrier 
with $L k_{F} = 4.0$ and $V_{0}/E_{F}=0.1$, for the couplings $(k_{F} a_{F})^{-1}$: $-0.8$ (dashed line); $0.0$ (dotted line); and $3.0$ (dash-dotted line).
The last curve is compared with the independent solution of the GP equation (full line).
Here $E_{F} = k_{F}^{2}/(2m)$ is the Fermi energy.
The inset shows the density profiles $n(x)$.} 
\end{center}
\end{figure}

Figures 1(a) and 1(b) present the spatial variation of the magnitude $|\Delta(x)|$ and phase $\phi(x)$ of the order parameter across a barrier centered at $x=0$, for 
three characteristic values of $(k_{F} a_{F})^{-1}$ and for the corresponding maximum value of the Josephson current.
The independent solution of the GP equation is also compared with the results of the B-dG equations for 
$(k_{F} a_{F})^{-1}=3$, when one expects the BEC limit to have been reached.
One sees that the depression of $|\Delta(x)|$ from its bulk value $|\Delta_{0}|$ due to the presence of the repulsive barrier causes a sharp variation of the phase 
$\phi(x)$ across the barrier in order to keep the current uniform, thus resulting in the total phase difference $\delta \phi$ accumulating across the barrier.
Both the depression of $|\Delta(x)|$ at $x=0$ and the value of $\delta \phi$ systematically increase from weak to strong coupling.
These features are reflected in the corresponding density profiles $n(x)$ shown in the inset of Fig.1(b).
In the BCS limit, Friedel oscillations modulated by $2 k_{F}$ affect $n(x)$ as well  as $|\Delta(x)|$ and $\phi(x)$.
\begin{figure}
\begin{center}
\epsfxsize=6.4cm
\epsfbox{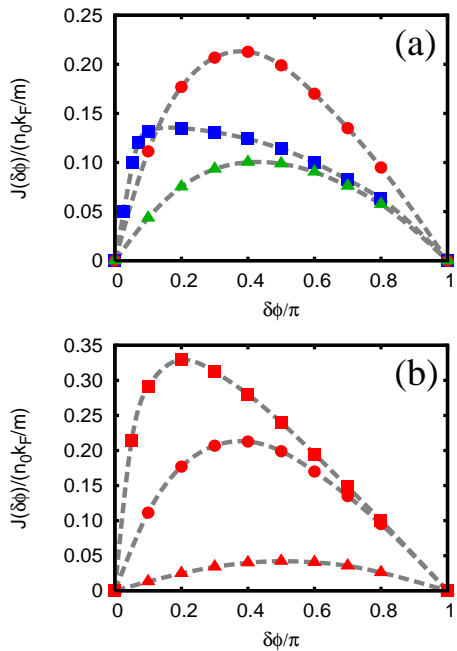}
\caption{(a) $J(\delta \phi)$ for the same values of $L$ and $V_{0}$ of Fig.1 and for the couplings $(k_{F} a_{F})^{-1}$:
$-0.8$ (squares); $0.0$ (circles); $1.0$ (triangles). 
(b) $J(\delta \phi)$ at unitarity for $L k_{F} = 4.0$ and different values of $V_{0}/E_{F}$:
$0.025$ (squares); $0.10$ (circles); $0.40$ (triangles).
[Dashed lines interpolate among data points.]} 
\end{center}
\end{figure}

The characteristic Josephson current/phase relation $J(\delta \phi)$ for a range of different coupling encompassing unitarity is shown in 
Fig.2(a), for the same values of $L$ and $V_{0}$ considered in Fig.1.
Note that  the Josephson current is considerably enhanced at unitarity with respect to the BCS and BEC sides, and that the curves $J(\delta \phi)$ stretch from being 
proportional to $\sin (\delta \phi)$ when $(k_{F} a_{F})^{-1} = 1.0$ to being (almost) proportional to $\cos (\delta \phi /2)$ when $(k_{F} a_{F})^{-1} = - 0.8$.
By scanning the coupling parameter for given barrier, we have further verified that: (i) In the BEC limit one invariably obtains 
$J(\delta \phi) \propto \sin (\delta \phi)$, reflecting the fact that in this limit the barrier heigth $V_{0}$ becomes larger than the relevant energy scale $\mu_{B}$; 
(ii) In the BCS limit (whereby the coherence lenght by far exceeds the barrier width) one reduces to considering a delta-like barrier with strength 
$Z = (Lk_{F}) (V_{0}/E_{F})$, for which $J(\delta \phi) \propto \sin (\delta \phi)$ when $Z\gg 1$ and $J(\delta \phi) \propto \cos (\delta \phi /2)$ when $Z \ll 1$.
Such a progressive evolution from a $\sin (\delta \phi)$-dependence for strong barriers to a $\cos (\delta \phi /2)$-dependence for weak barriers occurs actually 
for any coupling, as shown for instance at unitarity in Fig.2(b)~\cite{Sols}.

From these Josephson characteristics one can extract the maximum (critical) current $J_{c} = n_{0} q_{c}/m$ for given barrier and coupling.
The corresponding critical velocity $q_{c}/m$ is shown in Fig.3 vs $(k_{F} a_{F})^{-1}$ from weak to strong coupling and for several barrier heights.
All curves present a maximum at about unitarity, with the value of the maximum increasing as the ratio $V_{0}/E_{F}$ is decreased. 
The question thus naturally arises whether this limiting procedure of lowering the barrier 
height leads to an intrinsic upper value of $q_{c}/m$ for given coupling.

On physical grounds, a vanishingly small barrier acts as an impurity that probes the stability of the homogeneous superfluid flow, thus playing a similar 
role to the walls of the container in the context of the Landau criterion for superfluidity \cite{AGD}.
One then expects the velocities reported in Fig.3 to never exceed the corresponding value of the critical velocity obtained by the Landau criterion.
This critical velocity is determined by the available quasi-particle excitations which are of a different nature on the two sides of the crossover.
Specifically, on the BCS side there occur the pair-breaking excitations characteristic of BCS theory,
while on the BEC side one expects sound-mode quanta to be the relevant excitations.

Accordingly, in Fig.3 we have also plotted (full lines) the critical velocities corresponding to the above two branches of the Landau criterion.
In particular, the ``left'' branch corresponds to the expression $q_{c}^{2}/m = \sqrt{ \mu^{2} + \Delta_{0}^{2}} - \mu$, which is obtained by finding the smallest value of 
$q_c$ for which $E(k)-(q_c/m) k=0$ has solutions, where $E(k)=\sqrt{(k^2/(2 m) - \mu)^2 + \Delta_0^2}$ is the BCS single-particle energy of wave vector $k$ when the superfluid is at rest.   
The ``right'' branch corresponds instead to the sound velocity of the 
Bogoliubov-Anderson mode obtained within the BCS-RPA approximation \cite{MPS}, which generalizes the result $q_{c}^{2}/m = \mu_{B}$ valid in the BEC limit.
These two Landau branches represent an upper limit for the $q_{c}/m$ vs $(k_{F} a_{F})^{-1}$ curves obtained from our numerical calculations in 
the presence of a barrier, the latter consistently tending to the former from below as the barrier height is progressively decreased.
The two Landau branches intersect each other at about unitarity, where the maximum value of $q_{c}/m$ is then achieved~\cite{Randeria,Petrov}.

\begin{figure}
\begin{center}
\epsfxsize=6.5cm
\epsfbox{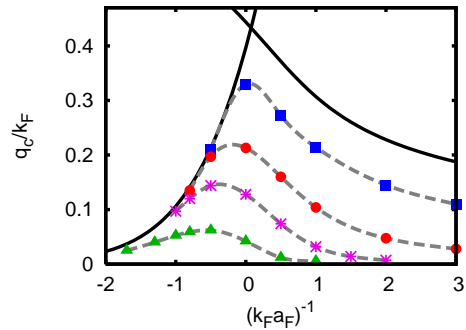}
\caption{Maximum velocity $q_{c}/m$ vs $(k_{F} a_{F})^{-1}$ for different barriers with $L k_{F} = 4.0$ and $V_{0}/E_{F}$:
$0.025$ (squares); $0.10$ (circles); $0.20$ (stars); $0.40$ (triangles).
[Dashed lines interpolate among data points.]
Full lines represent the two Landau branches for pair breaking (left) and sound mode (right).} 
\end{center}
\end{figure}

A few comments on how these two Landau branches result from the solution of the B-dG equations are in order.
As a consequence of Galilean invariance, no upper critical velocity would result from the equations (\ref{B-dG-equations}) and (\ref{self-consistency}) for a 
strictly homogeneous system.
It is thus \emph{only through a limiting procedure}, implemented, e.g., by a progressively vanishing barrier height, that an upper value 
of the critical velocity results when solving these equations.
Yet, by this procedure one would naively expect to recover only the left Landau branch corresponding to pair-breaking excitations, arguing that the B-dG equations 
correspond to the BCS mean field which contains explicitly only this type of excitations.
The fact that the right Landau branch for sound-mode excitations emerges, too, from our numerical calculations appears thus remarkable.
The crucial point is that, in the presence of a non-trivial geometry,  the imprint of the excitation spectrum is found already in the ground-state 
wave function at given (super) current, which in the present context corresponds to the mean-field equations (\ref{B-dG-equations}) and (\ref{self-consistency}) 
at zero temperature.
Since these equations reduce to the GP equation in the BEC limit, also their excitations on the BEC side correctly correspond to those of the 
GP equation~\cite{PS04}.  

The above results suggest us a procedure to extract the value of the pairing gap (order parameter) from the BCS limit to unitarity, which is  
the region of most interest for this quantity.
Previous studies~\cite{gap} have shown that for pair-breaking excitations the BCS expression $E(k)$ remains valid even beyond mean-field, with 
appropriate values of $\Delta_0$ and $\mu$. In this way, the Landau expression for $q_c$ reported above remains also valid, and can be inverted to give 
$\Delta_0/E_F= 2 (q_c/k_F) \sqrt{(q_c/k_F)^2 + \mu/E_F}$. The value of $\Delta_0/E_F$ is thus obtained from the experimental value
of $q_c/k_F$ (resulting, for instance, by extending to ultracold Fermi atoms the method recently used to study the flow of a bosonic condensate 
through a potential barrier~\cite{EA-2007}) and the value of $\mu/E_F$ determined either experimentally or from theoretical calculations. In this respect, 
we have verified that the differences in the values of $\Delta_0$ obtained by using mean-field or QMC values of $\mu$ never exceed 10\% over the 
relevant range of the coupling parameter.
In addition, the Landau value of $q_c$ is expected to be measured with reasonable accuracy on the BCS side of the crossover, where this limiting 
value is reached rather quickly by decreasing $V_0$ as shown in Fig.3.    
Measurement of $q_c$ would therefore provide a quantitative probe for fermionic superfluidity.
\acknowledgments
We are indebted to G. Deutscher, R. Hulet, and F. Pistolesi for discussions. 
This work was partially supported by the Italian MIUR under Contract Cofin-2005 ``Ultracold Fermi Gases and Optical Lattices''.



\end{document}